# Imaging structurally dynamic ribosomes with cryogenic electron microscopy

Samantha M. Webster[1], Mira B. May[1], Barrett M. Powell[1], and Joseph H. Davis[1,2,✉]

[1]Department of Biology, [2]Program in Computational and Systems Biology

Massachusetts Institute of Technology

Cambridge, MA 02139

✉ Correspondence: jhdavis@mit.edu

## ABSTRACT

Throughout the history of electron microscopy, ribosomes have served as an ideal subject for imaging and technological development, which in turn has driven our understanding of ribosomal biology. Here, we provide a historical perspective at the intersection of electron microscopy technology development and ribosome biology and reflect on how this technique has shed light on each stage of the life cycle of this dynamic macromolecular machine. With an emphasis on prokaryotic systems, we specifically describe how pairing cryo-EM with thoughtful experimental design, time-resolved techniques, and next-generation heterogeneous structural analysis has afforded insights into the modular nature of assembly, the roles of the many transient biogenesis and translation co-factors, and the subtle variations in structure and function between strains and species. This work concludes with a prospective outlook on the field, highlighting the pivotal role cryogenic electron tomography is playing in adding cellular context to our understanding of ribosomal life cycles, and noting how this exciting technology promises to bridge the gap between cellular and structural biology.

## Ribosomes and electron microscopy: a history intertwined.

The ribosome, which is responsible for protein translation in all domains of life, is a massive ribo-nucleoprotein complex ranging in size from ~2.5 MDa in bacteria[1] to ~4 MDa in eukaryotes[2]. This essential molecular machine is comprised of two subunits formed by interwoven RNA helices and ribosomal proteins that assemble through a dynamic, multi-step biogenesis pathway[3,4]. Once assembled, the machine is itself dynamic, undergoing structural changes coupled to its function in mRNA binding and decoding, peptidyl transferase activity, and its eventual sequestration and degradation. These dynamics involve both conformational motions and regulated compositional changes, with more than 50 core proteins and hundreds of transient binders and cofactors associating with the particle during its assembly and as it functions[5].

Electron microscopy (EM), including negative stain EM, single particle cryogenic EM, and cryogenic electron tomography, have proven to be uniquely suited for studying the ribosome. Unlike X-ray crystallography, in which a requisite crystal lattice locks the ribosome in a single structural state, cryoEM observes $10^4$-$10^7$ individual and, potentially, structurally heterogeneous particles in a single experiment. Although such structural heterogeneity presents computational challenges for the reconstruction and refinement of cryoEM density maps, it also provides an opportunity to see the full range of structural states populated in a sample[6]. With the advent of improved imaging technology[7,8] and 3D-reconstruction algorithms[9-11], ribosome structural dynamics are being systematically characterized in greater detail than ever before[12] (Fig. 1).

Early studies of the ribosome are inextricably linked with the development of electron microscopy. These "small particulate component[s] of the cytoplasm" that were briefly known as "particles of Palade," were first identified in 1955 in electron micrographs of the endoplasmic reticulum of rat liver cells[13,14], though Salvador Luria had incidentally imaged ribosomes in his electron micrographs of bacteriophages years prior[15]. Early evidence for co-transcriptional ribosome assembly, which is discussed below, came from direct visualization of rRNA synthesis in rapidly growing cells using "Miller spreads" – a method that gently disperses chromatin and fixes RNA polymerases to their DNA templates and nascent transcripts[16-18]. Negative stain electron micrographs of such Miller spreads revealed dense packing of RNA polymerases along *E. coli* rRNA operons, with fibrils of increasing lengths emanating from the RNA polymerase. These fibrils were decorated with protein-dense ribosome assembly intermediates, which collectively gave rise to a "Christmas tree" morphology (Fig. 2) that highlighted the co-transcriptional nature of ribosome assembly.

Such applications of electron microscopy not only advanced the field's understanding of ribosomes, but many were also critical in advancing electron microscopy as a technology. Indeed, the large, dense, and abundant nature of ribosomes make them nearly ideal imaging subjects[19] and, as such, they have played key roles in the development of negative stain and cryogenic EM[20,21], image processing and 3D-reconstruction methods[21-25], the development of early heterogeneous classification and reconstruction algorithms[26], and the more recent development of cryoEM and cryo-electron tomographic methods to determine structures *in situ*[27-30]. Likewise, some of the earliest examples of exhaustive structure classification methods aimed at understanding function[31,32], and the development of methods to model continuous forms of heterogeneity were tested using datasets of ribosome assembly intermediates that exhibited vast structural heterogeneity[6,11,12,33,34].

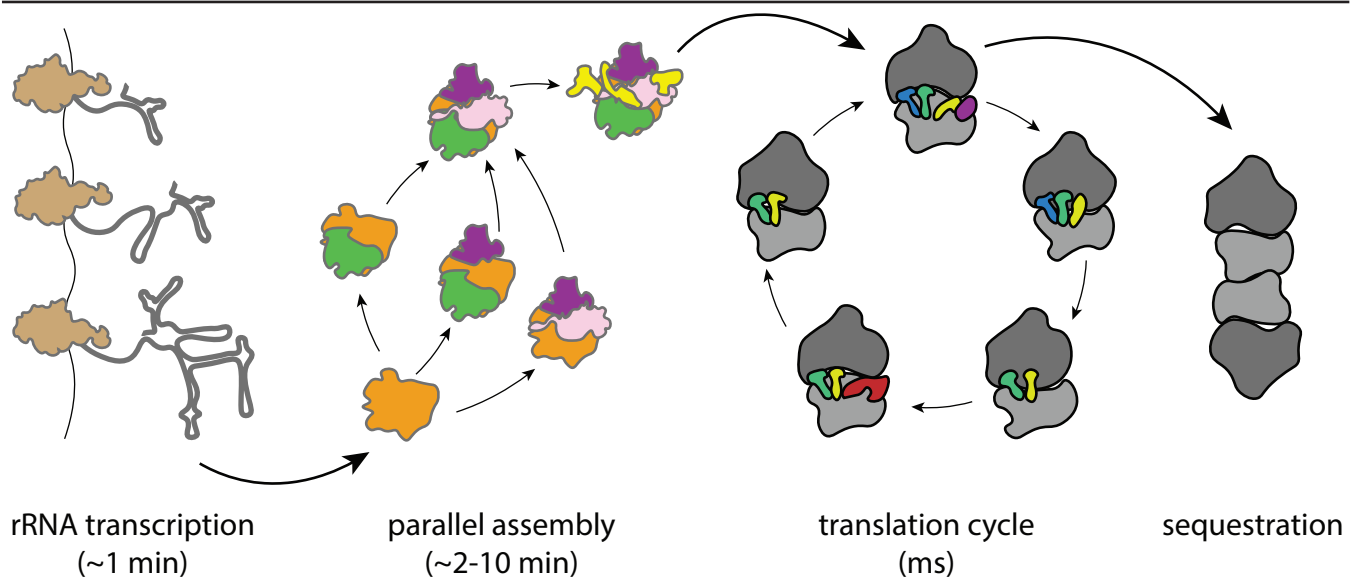

rRNA transcription (~1 min)  parallel assembly (~2-10 min)  translation cycle (ms)  sequestration

**Figure 1. Selected phases from the bacterial ribosome lifecycle.**
The ribosome lifecycle involves an array of heterogeneous structural states related by both parallel and sequential pathways across timescales spanning several orders of magnitude.

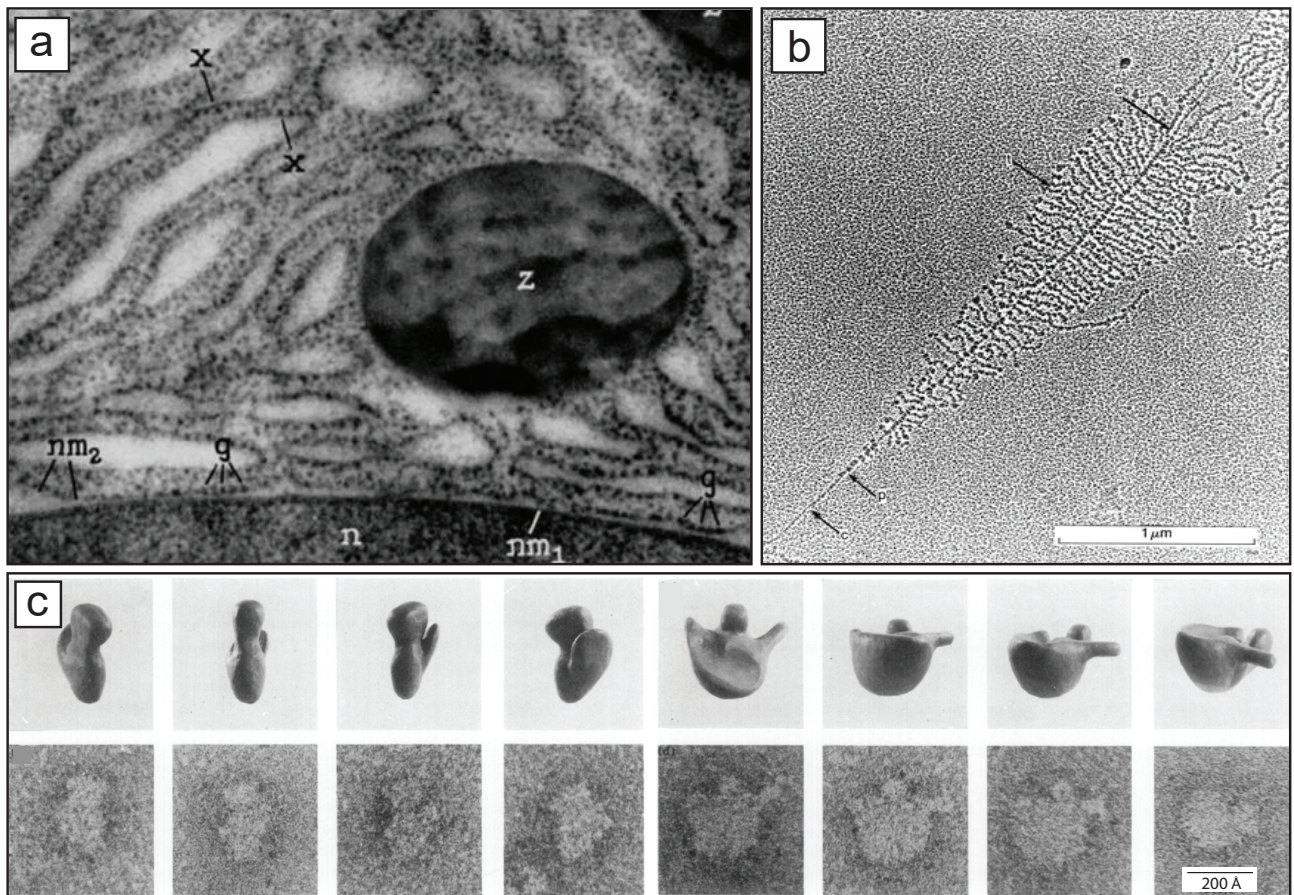

**Figure 2. Early studies of ribosomes and electron microscopy.**
**(a)** Electron micrograph of endoplasmic reticulum from rat pancreatic acinar cells, first published[13] by George Palade in 1954. g – Palade's "granules" i.e. ribosomes, n – nucleus, x – regions where granules appear to be attached to the membrane of the endoplasmic reticulum, $nm_1$ – nuclear membrane proper, $nm_2$ – endoplasmic reticulum membrane limiting the cytoplasm toward the nucleus, z – zymogen granule[13]. **(b)** Electron micrograph of a gene transcription unit prepared by the Miller spreading technique[158]. **(c)** The first 3-dimensional models of the 30S (left) and 50S (right) subunits, with single particle images from electron micrographs in corresponding orientations[20]. Figure panels reproduced with permission from above references.

In this chapter, we highlight how methodological advances in electron microscopy have enabled a greater understanding of the structural dynamics occurring in ribosomes as they undergo assembly and as they function, providing vignettes where electron microscopy has played a central role. Although EM has contributed to our understanding of ribosome structure-function relationships in both eukaryotes and prokaryotes, we focus on the latter, and direct readers to excellent recent reviews of analogous studies of eukaryotes[35,36].

### Snapshots of ribosome biogenesis.

Ribosome biogenesis is a rapid, multi-step process, involving the interplay of ribosomal proteins (r-proteins) and RNA (rRNA), as well as transiently bound assembly cofactor proteins. In prokaryotic cells, much of the particle's assembly occurs co-transcriptionally, with rRNA folding coupled to r-protein binding events that help to stabilize productively folded rRNA, thereby guiding the particle's maturation[4,18,37-39]. Notably, however, pioneering work by the Nomura and Nierhaus groups that reconstituted assembly of the 30S and 50S subunits *in vitro* showed that purified, fully transcribed rRNAs and r-proteins can self-assemble into active ribosomal subunits, which argued that co-transcriptional assembly is not strictly required for ribosome biogenesis[40-49]. Moreover, this *in vitro* approach allowed them to test the binding interdependence between r-proteins through order-of-addition experiments, and these studies provided evidence for thermodynamic coupling between many r-protein binding events. Certain proteins, termed primary binders, could bind the rRNA directly, whereas others, termed secondary and tertiary binders, exhibited stronger binding when other r-proteins were present. These experiments supported a model of hierarchical ribosome subunit assembly involving sequential binding of secondary and tertiary proteins, and parallel assembly of the primary binders.

Whereas these *in vitro* assembly reactions demonstrated the ability of subunits to assemble using only core ribosomal components, assembly is known to occur more efficiently and more rapidly in cells. Although the co-transcriptional nature of such cellular assembly is likely a dominant contributor to this improvement, a wide array of cellular assembly cofactors, including GTPases, rRNA and r-protein modification enzymes, and RNA helicases are also posited to contribute to the improved assembly efficacy observed in cells[4,38,50]. Additionally, many of these cofactors profoundly impact assembly efficacy under stress conditions, including cold-shock, suggesting roles for assembly cofactors in overcoming stress-related barriers to productive assembly[51]. Despite this broad understanding of assembly cofactor roles, there remain several areas of active inquiry, including: when in the assembly process do cofactors act; how do cofactors interact with each other; is there redundancy amongst cofactors; and how do cellular stresses impact the role of these cofactors? Moreover, basic questions about the mechanisms by which these proteins facilitate assembly remain unanswered, with postulated functions including: ensuring the correct assembly order[52-54]; expediting specific structural changes[55,56]; and proofreading specific assembly defects[12].

Answering these and related questions has proven challenging for three key reasons. First, despite its large size and complexity, the entire *E. coli* ribosome biogenesis process is completed in roughly two minutes and ribosomal precursors account for only 2-5% of all ribosomes during rapid growth[57,58]. As a result, isolating and studying specific ribosomal intermediates is difficult. Second, assembly cofactors are believed to associate transiently with their ribosomal substrates, and these substrates are of low abundance in the cell, making them difficult to isolate natively. Finally, multiple parallel assembly pathways exist that produce highly heterogeneous ensembles of assembly intermediates[3,31]. Approaches to isolate the relatively short-lived intermediates include: genetic systems to perturb r-protein and cofactor levels and thus stall the assembly process[31,53,59-62]; *in vitro* reconstitutions, which are inherently slow[63-65]; and biochemical approaches, such as affinity enrichment of assembly-factor bound particles[66]. Over the past 10 years, each of these strategies have been coupled to single particle cryoEM to characterize the structure and composition of highly heterogenous populations and, in some instances, to track dynamic structural re-arrangements that occur during assembly. Here, we highlight a non-exhaustive set of such studies that leveraged single particle cryoEM to understand the biogenesis process.

### CryoEM reveals conserved structural blocks in ribosome biogenesis.

Whereas the Nierhaus experiments provided a roadmap to understand the consequences of withholding an r-protein from an *in vitro* large subunit assembly reaction, it was unclear how cellular assembly pathways, which are under exquisite transcriptional and translational control, and bear an array of potentially buffering assembly cofactors, would respond to such a perturbation. To answer this question, Davis *et al.* depleted the early binding r-protein bL17 from cells, resulting in the accumulation of large subunit intermediates exhibiting both compositional and conformational heterogeneity. These intermediates were biochemically purified and analyzed using quantitative mass spectrometry and cryoEM[31] – a form of *ex vivo* analysis. In one of the first applications of deep structural classification to ribosome biogenesis intermediates, this work revealed 13 distinct intermediate structures resolved at 4-5 Å resolution that allowed for comparison of the presence/absence of each rRNA helix and r-protein (quantified as "occupancy") across the entire range of ribosomal precursors. This quantification produced an "occupancy map" that, upon hierarchically clustering, revealed sets of similar assembly intermediate particles and sets of rRNA helices and r-proteins whose occupancy profiles were correlated across the 13 structures. Guided by these occupancy maps, the structures were then arranged into an assembly pathway that exhibited hierarchical and parallel elements, and highlighted the malleability of bacterial ribosome assembly. For example, some r-proteins that were known to bind after bL17 in wild-type cells[67] were bound to these bL17-free particles, indicating that the assembly order could be permuted when limited quantities of an early binding r-protein were available. Moreover, this approach identified five major groups of rRNA and protein "blocks" that exhibited highly correlated behavior across the accumulating intermediates and co-localized on the mature large subunit. The authors suggested that this modular block-like assembly was a result of thermodynamic cooperativity built into the assembly process, and the subsequent observation of similar assembly blocks in independently analyzed large subunit assembly intermediates is consistent with this notion[3,64]. Finally, by identifying mutually exclusive blocks among ribosome precursors, which the authors posited would not disassemble once formed, this work provided the first structural evidence of parallel, branched assembly pathways (Fig. 3). Notably, more recent applications of advanced classification and reconstruction approaches, including deep hierarchical classification[32] and continuous models of structural heterogeneity have recapitulated these original analyses[6,11,33,34].

This study additionally allowed for expansion of the classic Nierhaus assembly map under perturbed *ex vivo* conditions to include rRNA helices and r-protein binding interdependencies, and provided insights into how different assembly pathways can permute the order and coupling of these r-protein binding and rRNA docking events. Taken together, the flexibility observed in the assembly pathways in this work further highlighted the value of interrogating assembly *in vivo*. Indeed, we and others have speculated that such flexibility may be an essential and evolutionarily selected feature of bacterial ribosome assembly as it would allow assembly to proceed in the face of transient shortages in particular r-proteins or assembly cofactors, or when cells encounter environmental conditions that disfavor a particular assembly pathway[3,68,69].

More recently, the Spahn and Nikolay groups have worked to build a structural understanding of the *in vitro* reconstituted assembly process pioneered by the Nierhaus group[63,64]. Importantly, these experiments detail progressive maturation pathways based only on the inherent biochemical properties of rRNAs and r-proteins, effectively insulating the observations from the complications that could arise from interactions with assembly cofactors or regulatory cascades in cells. Interestingly, the overarching themes observed by *ex vivo* Davis et al., including multiple parallel routes of assembly, were reinforced by these *in vitro* studies. Moreover, both the *in vitro* and *ex vivo*, ribosomal precursors appeared to mature in a domain-wise manner, with early precursors comprised of a stable core lacking a functional peptidyl transferase center and key intersubunit bridging rRNA helices, and a significant fraction of particles lacking a well resolved central protuberance. Of note, this reconstitution approach using purified components allowed Qin and colleagues to catalog one of the least mature assembly intermediates to-date, which may represent the minimal stably folded structural domain[64]. Interestingly, similar early assembly intermediates were recently discovered both in a sample isolated from an *in vitro* co-transcriptional assembly reaction[65] and that purified from cells lacking the helicase DeaD that were grown at low temperature[70], indicating such early assembly states can also be observed during co-transcriptional assembly *in vitro* and in particles isolated from cells.

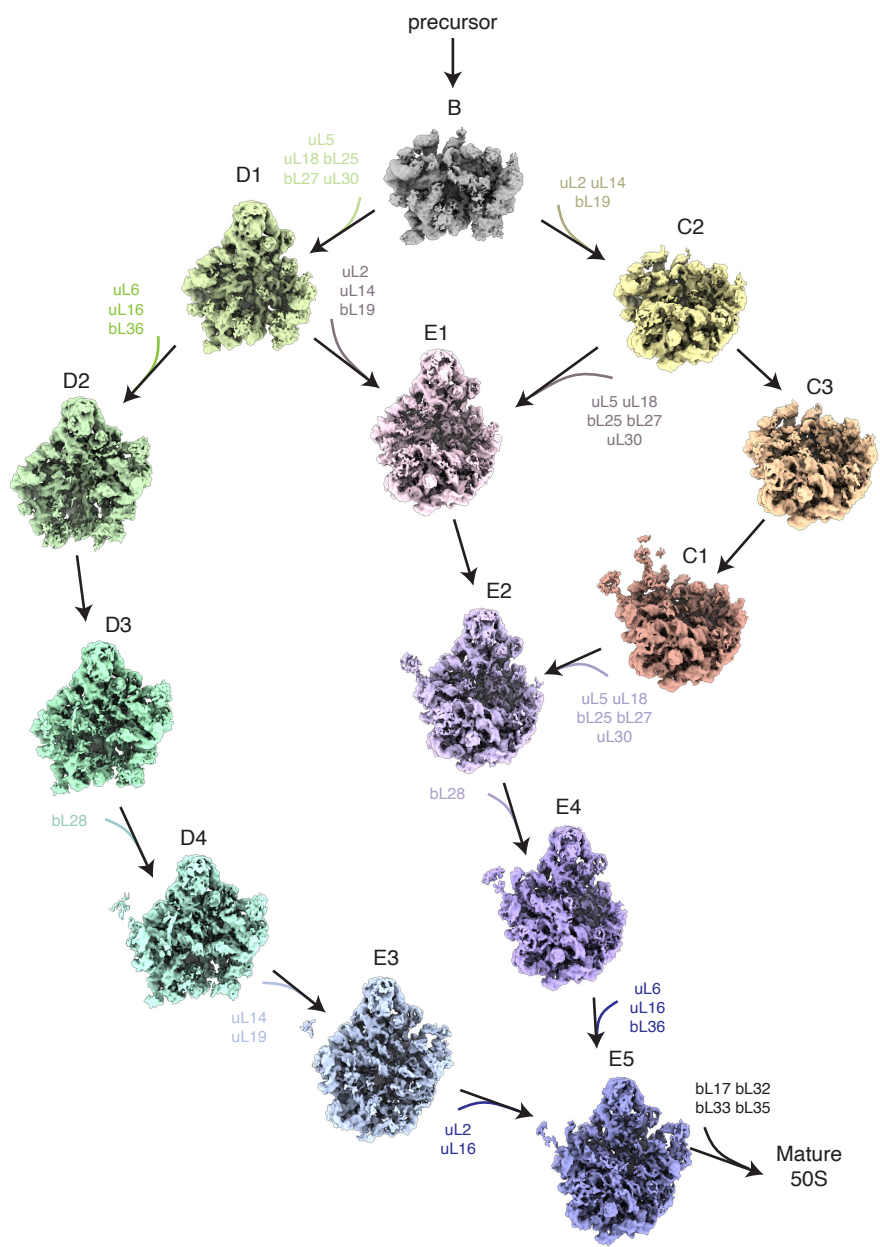

**Figure 3. Parallel assembly pathways observed under bL17 depletion.**
The most immature observed class B is shown at the top, which is hypothesized to transition to the most mature class E5 via several branched pathways (black arrows). Incorporated r-proteins are indicated at each transition step. Adapted from Davis *et. al* 2017[31].

**A new perspective on biogenesis cofactors.**
As above, cryoEM has significantly impacted our understanding of how ribosome biogenesis cofactors affect the assembly process, with most structural studies characterizing the ribosomal particles that accumulate in the absence of a cofactor of interest, or in determining the structure of the assembly-cofactor bound complex. To date, studies have investigated the 30S cofactors RsgA/YjeQ[71-76], RbfA[72,77], RimM[59,73], RimP[55], KsgA[12,78-80], and Era[81,82], and the 50S cofactors RbgA[52,53,60,83], YphC/EngB and YsxC[61], EngA[84], RrmJ[85], ObgE[66,86], SrmB[62], and DeaD[70]. Here, we highlight the biogenesis cofactors RbgA and KsgA, detailing how cryoEM and genetic perturbations to deplete cells of these cofactors have been applied to interrogate each cofactor's mechanism of action.

Notably, given the aforementioned flexibility and inherent permutability of the *in vivo* assembly pathways, interpreting such cofactor depletion experiments is challenging[3]. Indeed, parallel assembly branches raise the possibility that the accumulating particles one observes upon cofactor depletion are not true substrates for the cofactor but rather particles that have progressed towards a thermodynamically stable state that is no longer able to interact with the factor[3,87]. Indeed, it is even formally possible that such accumulated particles are not competent to mature, and are therefore by definition off-pathway.

In analyzing the particles that accumulate in the absence of the ribosome biogenesis-associated GTPase RbgA, Jomaa *et al*. considered these possibilities and, by coupling pulse-labeling to quantitative mass spectrometry, they determined that the majority of the RbgA-depletion particles were in fact competent to mature to functional 70S ribosomes[53,57]. Interestingly, they further noted that the functional core (*i.e.* the peptidyl transferase center, PTC) in these particles was highly disordered, despite observing that the majority of the other structural elements have adopted their mature conformations. Whereas the general observation that functional domains mature late predated this work, the effect was even more pronounced in these[53,60] and related[61] particles that accumulated upon co-factor depletion. This observation has been interpreted as a quality control mechanism inherent in the assembly process that prevents premature particles from engaging in translation until the particle is completely assembled[53] – akin to issuing homeowners keys only once the house is fully constructed, with assembly cofactors acting as building inspectors to ensure proper assembly order is maintained and particle maturation is complete before certification.

Another method for determining cofactor function is to add the protein of interest back to ribosomes purified from a cofactor-depleted strain and to then assess the changes that occur *in vitro*. Sun and colleagues employed this strategy while studying the function of KsgA[12], a methyltransferase which has long been suspected as a biogenesis cofactor involved in 30S subunit maturation[88,89]. The group characterized the ensemble of intermediates that accumulate in a cold-sensitive Δ*ksgA* strain of *E. coli* grown at low temperature, where these accumulating particles were presumed to be KsgA substrates. To assess the impact of KsgA addition to these substrates, they also analyzed structures resulting from addition of purified KsgA to this sample. To their surprise, the KsgA-treated structures were grossly *less* mature than the particles isolated from the depletion strain, with large swaths of poorly-resolved density in the head, platform and spur domains, indicating greater conformational or compositional heterogeneity in these regions. This led to a hypothesis that KsgA may recognize a structural defect in its substrates and induce disassembly of such particles. Using cryoDRGN[6,11] to analyze particles not yet treated with KsgA, they observed that a key helix (H44) could adopt two alternative conformations – one active and the other inactive. Quantitation of the relative abundance of these conformations revealed that upon treatment, KsgA preferentially depleted the sample of H44-inactive particles. Taken together, the evidence suggested that KsgA specifically recognizes inactive 30S intermediates and, through its binding, induces large-scale structural remodeling and disassembly that may afford these inactive particles an opportunity to re-fold and progress down a productive assembly pathway (Fig. 4). In sum, this recognition and disassembly activity was interpreted as KsgA-dependent kinetic proofreading[90] of the assembly process. In addition to illuminating a new proofreading role for KsgA, this work also highlighted the potential of next-generation cryoEM analysis tools such as cryoDRGN for analyzing large datasets bearing extensive structural variation, and in using these data to inform our understanding of structural signaling, remodeling, and dynamics.

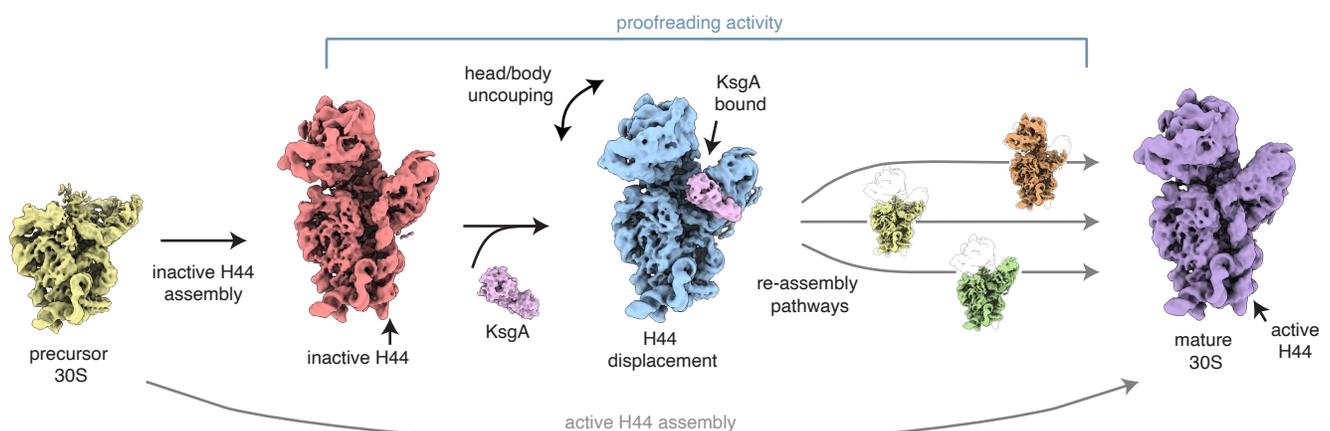

**Figure 4. KsgA proofreading model.**
During 30S assembly, immature particles with inactive H44 are recognized by KsgA, which induces subsequent uncoupling of the head/body to provide another opportunity for subunit maturation. Adapted from Sun and Kinman, *et al.* 2022[12].

Despite rapid recent progress in understanding bacterial ribosome biogenesis from a structural perspective, many key questions remain. For example, what is the relative flux through different assembly pathways; how do pathways change under different environmental conditions or cofactor depletions; and do multiple cofactors work in concert on a given particle, similar that observed in eukaryotic ribosome assembly intermediates? We expect that sophisticated biochemical and structural analysis protocols will be necessary to capture these more transient stages of assembly and, as techniques improve, we anticipate drawing deeper insights into both the immature assembly intermediates and the mature ribosome as it engages in its primary function in the cell — translation.

## A window into prokaryotic translation.

The structural dynamics of translation hold two major themes: compositional dynamics, with tRNAs and GTPase initiation and elongation factors binding and dissociating from the translating ribosome; and conformational dynamics, wherein the subunits undergo movements linked to function[91,92]. For example, during translation elongation, the large (LSU) and small (SSU) subunits rock back and forth relative to each other in a distinct ratcheting motion that is critical to translocating tRNAs through the active sites[93-95], as well as a subsequent step where the SSU head rotates relative to its body in order to move along the mRNA by one codon[96].

Due to limitations in microscope and detector hardware, early cryoEM structures of translating ribosomes were relatively low resolution[94,97-99]. Despite this, targeted studies based on previous biochemical and biophysical work[100-103] assembled desired complexes *in vitro* or stalled the translation complex pharmacologically or genetically, resulting in structures that significantly advanced our understanding of the mechanism and dynamics of translation. Such approaches provided the first visualization of ribosome-bound EF-G, a GTPase that catalyzes mRNA and tRNA translocation during polypeptide elongation[104]; facilitated discovery of the ratchet motion between the small and large subunits that occurs upon binding of EF-G[95]; and provided structures of tRNA in hybrid state between the P- and E-sites that occurs as a prelude to tRNA translocation[105]. Early structures also complemented crystallographic approaches and led to a general understanding that the factors EF-Tu, EF-G, RF-3, and IF-2 bind in the flexible GTPase Associated Center (GAP) via an "induced fit" model, wherein each subsequent factor acts to stabilize the ribosome and promote productive conformational changes to advance the translation cycle[97,104,106-108].

Over the past 15 years, technical advances in cryoEM[7,109] have produced ever-higher resolution structures that led to new insights into ribosome structure and function. For example, the sub-3 Å resolution structure of a ribosome•EF-Tu complex bound to aminoacyl-tRNA and the antibiotic kirromycin resolved all 35 rRNA modifications[110], providing a structural framework to understand antibiotic action and resistance[110-112], and it paved the way for a recent study determining 17 high-resolution structures of antibiotic-ribosome co-complexes[113]. Further methodological improvements culminated in a 1.6 Å resolution structure of a translating ribosome that, incredibly, allowed for direct structure-guided sequencing of the rRNA[114]. These structures additionally visualized the flexible and enigmatic protein bL9 in a closed conformation and identified residues on bL9 (Glu87) and uS6 (Arg24) that stabilized this closed conformation via a salt bridge. Notably, bL9 is suspected to play a role in fidelity of translation[115], frameshifting[116,117], polysome formation[118], and quality control pathways in collided ribosomes[119,120], and the exchange between the extended conformation originally observed in crystal structures at a lattice interface[121-123] and the closed conformations observed by Fischer *et al.* and others may help to explain these multi-modal functions.

Similarly, image analysis tools aimed at handling structural heterogeneity have begun to link mechanism to the structural ensembles within cryoEM datasets that can be computationally resolved. Loveland *et al.* used such approaches to resolve six near-atomic-resolution ribosome•tRNA•EF-Tu decoding complexes, bearing cognate and near-cognate tRNA-mRNA pairs[124]. These structures established that discrimination between cognate and near-cognate tRNAs hinges on a single base on the shoulder of the 30S subunit (G530), which can assume an "ON" conformation when properly stabilized by a network of hydrogen bonds between the mRNA and tRNA bases. They further showed that movement into the "ON" conformation aids in EF-Tu activation that precipitates GTP hydrolysis and EF-Tu dissociation from the complex, which represents the final step of aminoacyl-tRNA accommodation into the LSU tRNA A-site. The ability to visualize several related complexes in such detail was critical in elucidating the mechanisms of translational fidelity, and highlighted the value of accurate particle classification algorithms to not only improve resolution, but to also reveal the structural underpinnings of central biological phenomenon.

Whereas the combination of biochemical data and the targeted enrichment of structural states using genetic or pharmacological approaches has allowed researchers to infer the sequence of structural progression[3], the field has long-sought to directly determine time-dependent structural ensembles[125] and to visualize transitory structural states within these datasets. Pursuing these goals, Carbone et al. analyzed translational elongation as a function of time after the addition of EF-G using standard vitrification equipment[126]. Despite the relatively poor temporal resolution afforded by such plunge-freezing post-mixing, their intermediate timepoints captured transient translation states with EF-G bound, and a comparison of their structures highlighted the progression of structural states during EF-G-mediated translocation (Fig. 5). They noted that EF-G's extended conformation both pre-translocation (Structure III) and nearly post-translocation (Structure IV) was consistent with a model of EF-G acting as a rigid pawl, as opposed to a flexible motor that is actively driving translocation through GTP hydrolysis. They concluded that the motion driving translocation, therefore, is inherent and spontaneous internal movement of the ribosome, including the 30S body rotation and head swivel. The study also captured a structure (Structure VI) that did not have EF-G bound, but which was exclusively present in t=24 seconds (s) dataset, indicating that it was a transient state formed directly after EF-G dissociation. The structure showed the 30S head still in the swiveled position, indicating that the reversal of the head swivel occurs after EF-G dissociation.

Building on the promise of time-resolved cryoEM, recent innovations including the development of fast-mixing microfluidic vitrification instruments[127,128] have enabled millisecond-resolved studies of translation initiation and termination[129,130]. Kaledhonkar et al. used a rapid mixing approach to study the later steps in bacterial translation initiation, mixing 30S Initiation Complexes (IC) with 50S particles to investigate mechanisms of 70S IC formation and its subsequent maturation into the 70S Elongation Complex (EC). Using a custom-built microfluidic device for mixing and rapid vitrification, they collected datasets spanning 20-600 milliseconds (ms) post-mixing. They found that the 70S IC complex abundance peaked near 80 ms, representing ~40% of the particles, and that by the end of the 600 ms more than 60% of the particles were fully converted to 70S EC complexes (Fig. 5a). Notably, this rapid mixing approach was vital in resolving the short-lived 70 IC complex in the absence of perturbations, and helped to distinguish on-pathway complexes from off-pathway intermediates that could result from such perturbations. Inspection of this native 70S IC structure revealed that IF-1 had dissociated, which alleviated a steric clash with H69 of the large subunit and allowed the formation of inter-subunit bridges required for translation. Moreover, in the transition from the 70S IC to the 70S EC, the authors observed dissociation of IF-2, which appeared to trigger a reverse rotation of the small subunit by 3Å, thereby stabilizing the non-rotated inter-subunit orientation. Finally, the combination of this rotation and the absence of IF-2 enabled the fMet-tRNA to move from the peptidyl/initiation (P/I) configuration observed in the 70S IC to the peptidyl-peptidyl (P/P) configuration observed in the 70S EC, effectively placing the fMet moiety in a peptidyl-transfer competent position within the P-site, and enabling the complex to proceed with the first round of translation elongation.

At the other end of translation, Fu et al. used a similar technique to study the structural changes in release factors (RFs) during translation termination[130]. RFs contain structural motifs involved in recognizing a stop codon, as well as a GGQ motif that is responsible for ester bond hydrolysis, which allows the nascent peptide release from the ribosome. Interestingly, crystal structures of isolated bacterial RFs are compact, with only 20 Å between the stop codon recognition and GGQ motifs, whereas ribosome-bound RFs adopt an extended conformation that spans the ~70 Å between the decoding center and the peptidyl transferase center. If the compact conformation observed crystallographically is physiologically relevant, this implies that either: 1) the extended and compact conformations are in equilibrium in solution and the ribosome captures the extended form; or 2) the compact conformation binds and then extends on the ribosome. To distinguish between these possibilities, Fu *et al.* used a rapid mixing technique paired with cryoEM to study the structural progression after mixing a UAA-programmed release complex ($RC_0$) with RF-1. At 24 ms, 25% of ribosome-bound RF was in its compact form, in a sort of pre-accommodation state previously observed for RF-2 by cryo-EM in ribosomes stalled at the terminus of an mRNA lacking a stop codon[131]. This population rapidly decayed by 60 ms, into a ribosome-bound extended conformation RF complex (Fig. 5c), indicating 1) that RF-1 can bind in a compact state; and 2) that the conformation shift to the elongated state happens very rapidly upon RF binding, whereas the hydrolysis step, which acts to release the nascent polypeptide chain, occurs on a much longer time scale. The authors hypothesize that the compact form of RF may assist with rapid factor binding and more accurate stop-codon decoding.

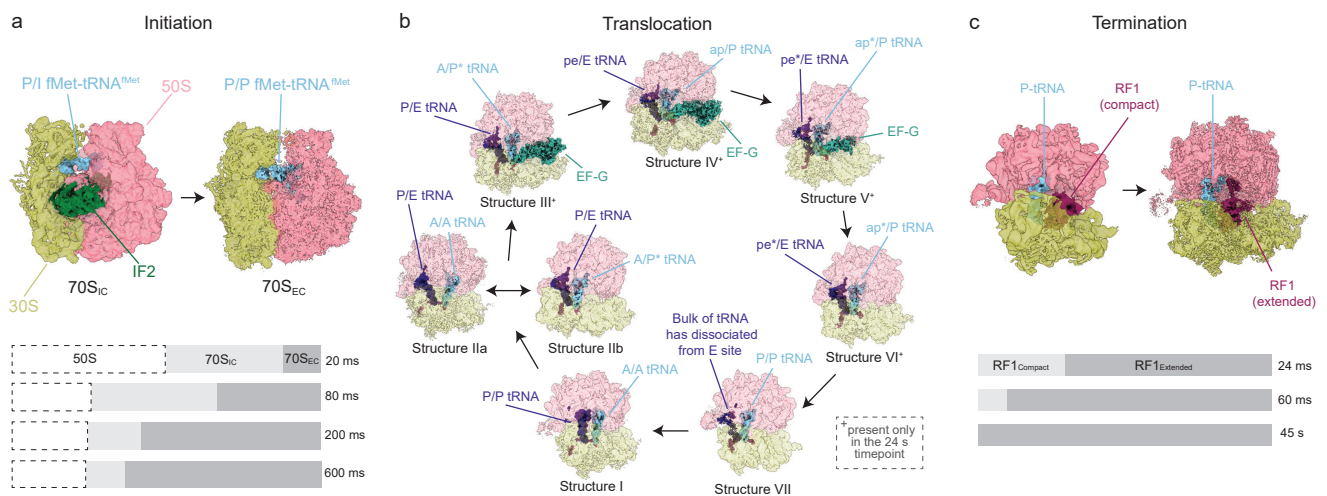

**Figure 5. Time-resolved cryoEM experiments reveal new structures at different steps of translation.**
(a) 30S Initiation Complex (IC) particles were combined with 50S particles using a mixing-spraying time resolved cryoEM set-up, resulting in four timepoints spanning 0-600 milliseconds (ms). The population of $70S_{IC}$ particles initially increased as the 50S particles bound the $30_{IC}$ particles, before the $70S_{IC}$ particles converted to 70S Elongation Complexes ($70S_{EC}$). Adapted from Kaledhonkar et al. 2019[129]. (b) EF-G was mixed with pre-translation 70S ribosomes loaded with tRNA$^{fMet}$ in the P site and dipeptidyl-tRNA$^{Pro}$ in the A site. Grids were blotted at 0, 24, and 600 seconds (s). The structures show a full progression of the translocation cycle, and include many tRNA hybrid states. Structures III, IV, V, and VI were found only in the 24 s sample. Structure V shows EF-G after domains I and II have partially dissociated. Adapted from Carbone et al. 2021[126]. (c) Using a rapid-mixing cryoEM set-up, UAA-programmed release complex, ($RC_0$) with tripeptidyl-tRNA in the P site was added to RF1. At 24 ms, roughly 30% of complexes were bound to RF1 in its compact form, while at 60 ms more than 90% of RF1 had converted to its extended conformation. Adapted from Fu et al. 2019[130].

### Responding to stress: a glimpse at sequestration.

Nutrient starvation and other stress can lead to translational suppression, which directs cellular resources away from the energy-intensive process of protein production and thereby promotes cell survival[132]. In *E. coli*, hibernation factors work in concert to suppress protein synthesis by binding to and blocking functional sites on the SSU, and by inducing the reversible dimerization of 70S ribosomes into a translationally-inactive 100S sequestration particle[133,134]. In contrast to ribosome degradation, which has been observed in mammalian cells[135], the generation of such a particle effectively allows cells to downregulate translation during starvation, and to then rapidly reinitiate translation should nutrients become available. The first structures of this sequestration particle were generated through cryoEM, and the visualization of these particles dimerized head-to-head through the SSU provided a structural understanding of translational repression in bacteria[136-138]. Interestingly, although hibernation factors vary between species, the process of sequestration may be conserved as a 110S sequestration particle has been identified in mammalian cells[139].

### What lies ahead.

Whereas single particle analysis of purified ribosomes has provided key mechanistic insights into ribosome biology, there are a number of benefits to studying samples *in situ* – that is, in their native cellular environment. By avoiding cell lysis and downstream purification, one can interrogate minimally perturbed ribosomes that are likely to better retain transiently bound cofactors. Moreover, *in situ* imaging can reveal sub-cellular localization patterns of sub-nanometer scale features that may be associated with biological function.

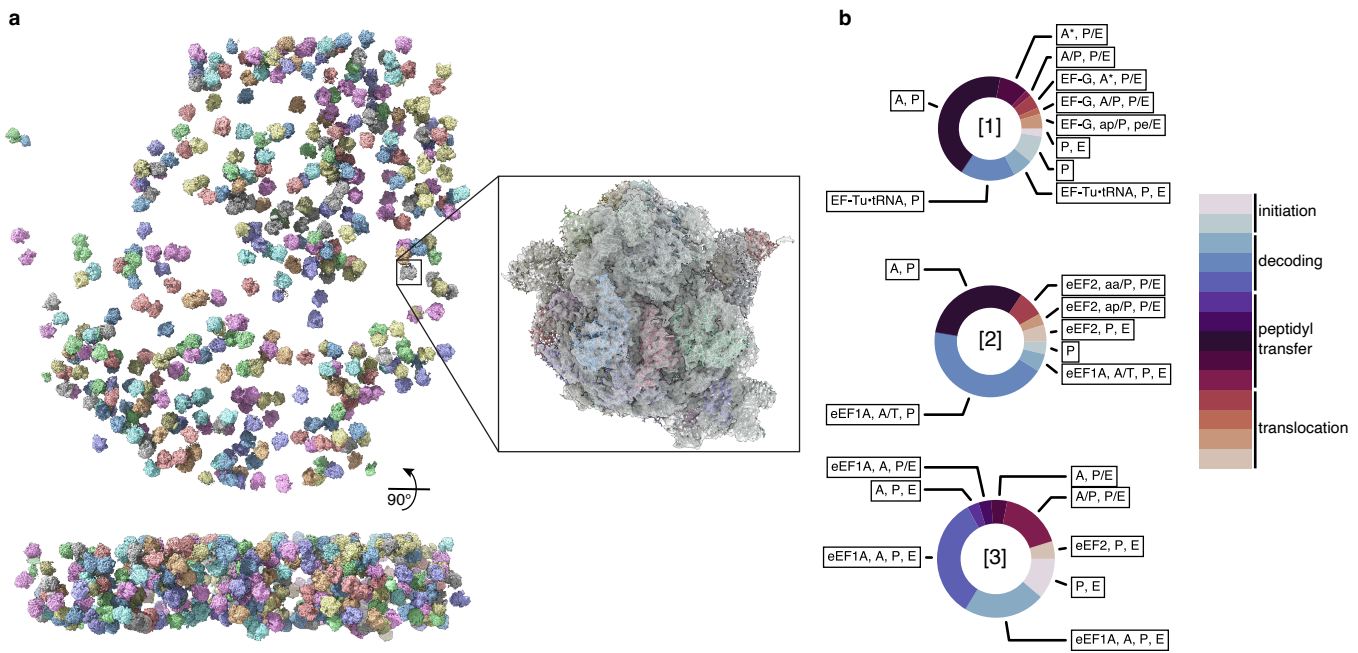

**Figure 6. Cryogenic electron tomography (cryoET) enables new insights into translation *in situ*.**
(**a**) Illustrative example of distinct ribosome translational states identified by cryoET sub-tomogram averaging and classification, mapped back to the cell and colored by translation state. Inset: Example of near-atomic resolution maps (gray, semi-transparent) from cryoET *in situ*, with fit atomic model pdb:7PHB (ribbon multi-color). Data were reprocessed from Xue *et al.*, 2022[118]. (**b**) Distribution of translation cycle states calculated from classification results of three recent *in situ* studies, highlighting similarities in relative abundance of states associated with various stages of the translation cycle observed by Xue *et al.* 2022[118] [1], Hoffman *et al.* 2022[159] [2], and Gemmer *et al.* 2023[160] [3].

Studying ribosomes *in situ* via electron microscopy is typically accomplished using cryo-electron tomography (cryoET), often preceded by focused ion beam (FIB)-milling, which is used to thin a cellular sample to allow transmission of elastically scattered electrons. In contrast to cryoEM, cryoET repeatedly images the same ~50-300 nm thick sample at different stage tilt angles, which helps to disentangle particles at different depths along any one projection axis. Such a "tilt series" can then be back-projected to reconstruct a 3D tomogram of the field of view typically 0.05-0.5 µm$^3$, thereby visualizing a significant portion of small bacterial cells, and a relatively small subregion of typical eukaryotic cells. From this low signal-to-noise ratio tomogram, ribosomes can be identified, extracted, aligned, and averaged to increase the signal-to-noise ratio and structural resolution in a process called sub-tomogram averaging (STA)[140].

Until recently, cryoET and STA structures rarely exceeded ~1-2 nm resolution, which limited the field's ability to glean molecular insights from the resulting structures. However, recent advancements in hardware for FIB-milling and imaging[141,142], strategies for data collection[143-145], and image processing software[28,146-153] have poised structural biology for an "*in situ* resolution revolution" akin to the single particle analysis "resolution revolution" of the early 2010s[7]. Downstream tools to analyze structural heterogeneity have also expanded, allowing identification of rare and dynamic subpopulations of ribosomes *in situ*[30,151,154-157]. Taken together, the field now enjoys faster and easier data collections, larger and higher quality datasets, more powerful data processing, and novel data analysis opportunities.

While many of these developments are recent and their synergistic effects are likely to bear fruit over the next several years, several exciting findings have already hinted at what is to come (Fig. 6). Tegunov *et al.* demonstrated that STA could resolve chloramphenicol-treated *M. pneumonaie* ribosomes at 3.5Å resolution *in situ*[28], and Xue *et al.* further analyzed the same system to characterize the distribution of translational states under antibiotic and untreated conditions *in situ*, resolving 13 compositionally and conformationally distinct states that recapitulate the known translation cycle[118]. Notably, detailed analysis of native polysomes *in situ* allowed Xue *et al.* to directly observe protein bL9 playing a role in polysome coordination and collision avoidance. In a novel approach to *in situ* imaging and processing, Lucas *et al.* designed 2D template matching (2DTM) which searches individual 2D micrographs for instances of a high-resolution template[29]. In one application of 2DTM they used a *M. pneumonaie* 50S subunit structure as a "bait", identifying via local searches the variable presence of adjacent smaller 30S subunits. In principle this approach could enable quantitation of small subpopulations of ribosomes with a particular cofactor of interest bound. Finally, we have developed a deep-learning based approach to study structural heterogeneity among subtomograms and applied it to the *M. pneumonaie* dataset first reported by Tegunov *et al*. In addition to recapitulating many of the aforementioned insights, this tool additionally resolved a minor population (~2%) of ribosomes associated with the membrane, thereby simultaneously visualizing a moderate resolution structure of a 70S ribosome - SecYEG holotranslocon co-complex *in situ*[30].

With this rich toolkit to pursue structural studies *in vitro* and *in situ*, we anticipate a wave of new biologically insightful findings in the coming years. For example, one can now directly develop and test structure-function hypotheses in the native cellular environment, and subsequently elucidate the specific structural mechanism involved with ~10$^{-2}$ second time resolution *in vitro*, all at near-atomic resolution. Such tools will be critical to address key remaining questions in the field, including: do ribosome biogenesis, translation, and sequestration, along with their associated cofactors, exhibit specific subcellular distributions; are such distributions perturbed in diseased states; and do the structural states captured *in vitro* accurately reflect states and their relative populations in the cell? We also posit that *in situ* approaches are likely to enable discovery of novel or weakly interacting cofactors not captured with traditional cryoEM using purified samples. We eagerly await the insights that lie ahead for the study of ribosome dynamics at the intersection of structural, molecular, and cellular biology enabled by traditional cryoEM, time-resolved cryoEM, and cryoET.

### Acknowledgments.
This work was supported by NIH grants R01-GM144542, 5T32-GM007287, and NSF-CAREER grant 2046778. Research in the Davis lab is supported by the Whitehead Family and the Sloan Foundation.